\documentstyle[aaspp4,flushrt]{article}

\begin{document}

\title{Spectral Hardness Decay with Respect to Fluence \\
in BATSE Gamma-Ray Bursts \\
}

\author{A. Crider and  E. P. Liang}
\affil{Department of Space Physics and Astronomy, 6100 S. Main,
 	Rice University, Houston, TX 77005-1892}

\author{R. D. Preece, M. S. Briggs, G. N. Pendleton, and W. S. Paciesas}
\affil{Department of Physics, University of Alabama in Huntsville,
 	Huntsville, AL  35899}

\and

\author{D. L. Band and J. L. Matteson}
\affil{Center for Astrophysics and Space Sciences 0424, University of California
	at San Diego, La Jolla, CA 92093}


\newcommand{\Epk}{\rm{E}_{\rm{pk}}}

\newcommand{\TotalBursts}{126}
\newcommand{\DiscardedBursts}{100}
\newcommand{\SelectedBursts}{26}

\newcommand{\BandBursts}{26}
\newcommand{\BrokBursts}{0}
\newcommand{\CompBursts}{0}

\newcommand{\CountPercent}{70\%}

\newcommand{\TotalPulses}{41}
\newcommand{\SelectedPulses}{41}
\newcommand{\GoodPulses}{41}
\newcommand{\BadPulses}{0}
\newcommand{\PercentagePulsesConsistent}{100\%}

\newcommand{\GoodEdPulses}{35}

\newcommand{\AverageLogPhi}{1.75 \pm 0.07}
\newcommand{\FWHMLogPhi}{1.0 \pm 0.1}

\newcommand{\DfromQfig}{0.17}
\newcommand{\PfromQfig}{0.18}

\newcommand{\PminX}{0.84}
\newcommand{\PsumX}{0.49}
\newcommand{\PprodX}{0.90}

\abstract{We have analyzed the evolution of the spectral hardness parameter, 
$\Epk$ (the maximum of the 
$\nu F_{\nu}$ spectrum) as a function of fluence in gamma-ray bursts.
We fit $\SelectedPulses$ pulses within $\SelectedBursts$ bursts with the
trend reported by Liang \& Kargatis (1996)
which found that $\Epk$ decays exponentially with
respect to photon fluence $\Phi \rm{(t)}$.  We also fit these pulses
with a slight modification of this trend, where $\Epk$ decays
linearly with energy fluence.  In both cases, we found the
set of $\SelectedPulses$ pulses 
to be consistent with the trend.  For the latter trend, which we believe
to be more physical,
the distribution of the decay constant
$\Phi_0$ is roughly log-normal, where the mean of
$\rm{log_{10}\Phi_{0}}$ is $\AverageLogPhi$ and the FWHM of 
$\rm{log_{10}\Phi_{0}}$ 
is $\FWHMLogPhi$.  
Regarding an earlier reported invariance in $\Phi_{0}$ 
among different pulses in a single burst, we found probabilities 
of $\PsumX$ to $\PminX$ (depending on the test used)
that such invariance would occur by coincidence, 
most likely due to the narrow distribution of $\Phi_{0}$ values 
among pulses. 
%
}


\section{Introduction}

The discovery of a 
gamma-ray burst optical counterpart with a measurable redshift
seems to have shown that the sources are cosmological in origin
(\cite{djor97,metz97}).
The observed fading multi-wavelength afterglows are so far consistent 
with the simple relativistic blastwave model (\cite{mesz97})
which radiates via a synchrotron shock.  However, the emission mechanism
resulting in the prompt gamma rays remains a mystery.
Studies of gamma-ray burst (GRB) spectral evolution have 
uncovered many trends which may be used to test possible emission
mechanisms.  In general, studies of GRB spectral evolution have 
focused on the ``hardness'' of bursts, measured either by the  
ratio between two detector channels or with 
more physical variables such as the spectral break
or peak power energy $\Epk$ (\cite{ford95})
which is the maximum of $\nu F_{\nu}$, where
$\nu$ is photon energy and $F_{\nu}$ is the specific energy flux.
Such hardness parameters were found to either follow a 
``hard-to-soft'' trend (\cite{norr86}),
decreasing monotonically while the flux rises and falls,
or to ``track'' the flux during GRB pulses (\cite{gole83,karg94}).  

The discovery that $\Epk$ often decays exponentially in bright, 
long, smooth BATSE GRB pulses 
\textit{as a function of photon fluence} $\Phi$
($=\int_{t'=0}^{t'=t} \rm{F_{N}}(t') dt'$)
(\cite{lian96}, hereafter LK96) provided a new constraint for emission models 
(\cite{lian97a,lian97b,daig98}).
In their analysis, LK96 fit the function
\begin{equation}
\Epk(t) = \rm{E}_{\rm{pk(0)}} \rm{e}^{-\Phi(t) / \Phi_{0}^{\rm{LK}}} 
\end{equation}
to 37 GRB pulses in 34 bursts.  
To interpret this empirical trend, they differentiated
Eq. 1 to find 
\begin{equation}
-\rm{d}\Epk/\rm{dt} = \Epk~\rm{F}_{N} / \Phi_{0}^{\rm{LK}} \approx \rm{F}_{E} 
/ \Phi_{0}^{\rm{LK}}
\end{equation}
where $\rm{F}_{E} = \int_{\rm{E} \approx 30 
\rm{keV}}^{\rm{E} \approx 2000 \rm{keV}} 
\rm{E}~N(\rm{E})~\rm{dE}$ is the BATSE energy flux 
(see Eq. 1 of LK96).  In this paper, we wished to avoid the
assumption that $\Epk~\rm{F}_{N} \approx \rm{F}_{E}$.  To do this, we directly 
tested the trend $-\rm{d}(\Epk)/\rm{dt} = \rm{F}_{E} / \Phi_{0}$ 
by integrating it to give us the function
\begin{equation}
\Epk(t) = \rm{E}_{pk(0)} - \mathcal{E}\mathrm{(t)} \mathrm{/ \Phi_{0}}
\end{equation}
where $\mathcal{E}\mathrm{(t)}$ 
($=\int_{t'=0}^{t'=t} \rm{F_{E}}(t') dt'$) is the BATSE energy fluence. 
We emphasize that this is not a fundamentally different trend from the form
used in LK96.  

The decay constant $\Phi_{0}^{\rm{LK}}$ 
appeared to be invariant among pulses during some bursts
analyzed in LK96, suggesting
that individual pulses in a burst may originate in the same plasma. 
These discoveries coupled with the observed evolution of the spectral shape 
(\cite{crid98a}) suggested that saturated
inverse Comptonization may be a viable mechanism during the
gamma-ray active phase of bursts (\cite{lian97a}),  
regardless of the distance scale (\cite{lian97b}).   


\section{Procedures}

To determine the evolution of GRB spectral shapes,
we examined High Energy Resolution
data collected from the BATSE Large-Area Detectors (LADs) 
and Spectroscopy Detectors (SDs) on board the Compton
Gamma-Ray Observatory (\cite{fish89}). 
We began with the $\TotalBursts$ bursts
which appear in \cite{pree98}.
These bursts were chosen for having a BATSE 
fluence (28-1800 keV) $> 4 \times 10^{-5}$ erg cm$^{-2}$
or a peak flux (50-300 keV on a 256-ms time scale) 
$> 10~\rm{photon~s^{-1}~cm^{-2}}$.
The counts from the detector most nearly 
normal to the line of sight of each burst (burst angle closest to 0) were
background-subtracted and binned into  
time intervals each with a SNR of $\sim 45$ within the 28 keV to 1800 keV 
range.  Such a SNR has been found to be necessary in time-resolved
spectroscopy of BATSE gamma-ray bursts (\cite{pree98}).

We deconvolved the gamma-ray spectra of each time interval using 
the Band et al. (1993) GRB function

\vspace{4mm}

\( \begin{array}{rclrr}
\rm{N_{E}(E)} & = & \rm{A\left(\frac{E}{100~\rm{keV}}\right)^{\alpha}
   exp\left(-\frac{E}{E_{\rm{0}}}\right),} & 
   \rm{(\alpha-\beta)E_{\rm{0}} \geq E,} \\
\mbox{} & = & \rm{A\left[\frac{(\alpha-\beta)E_{\rm{0}}}{100~
\rm{keV}}\right]^{\alpha-
   \beta} exp(\beta-\alpha) \left(\frac{E}{\rm{100~keV}}\right)^{\beta},}
& \rm{(\alpha-\beta)E_{\rm{0}} \leq E,} & (4)
\end{array} \)

\stepcounter{equation}

\vspace{4mm}

\noindent where A is the amplitude (in $\rm{photon}~
\rm{s}^{-1}~\rm{cm}^{-2}~
\rm{kev}^{-1}$) and $\rm{E}_{0} = \Epk / (2 + \alpha)$. 
While LK96 assumed that $\alpha$ and $\beta$ were constant
during the course of each burst, 
this has since been shown to be untrue with a larger data set. 
(\cite{crid97}).
We thus left $\alpha$ and $\beta$ as free parameters in our fits.

At this point, we needed to select pulses within our bursts that we could 
use to test Equations 1 and 3.  Ideally, our pulses would not overlap other
pulses and our method for choosing the time bins associated with a pulse would 
not be biased. 
Unfortunately, 
by forcing our time bins to have a SNR $\sim45$ so that spectra may be fit to them, 
much time resolution is lost.  Pulses which would be easily separable at a higher time 
resolution become blurred together.  
In Figure \ref{Figure1}, we show an example of what 
would likely be identified as two pulses
in our coarse (SNR $\sim45$) data. 
Below it, 64-ms count rate data 
for this same burst, obtained
from the Compton Observatory Science Support Center (COSSC),
is plotted.  With higher time resolution, 
we see this bursts is composed of at least 4 distinct pulses.

To avoid contaminating our sample with overlapping pulses (and to avoid biases introduced by
a human in pulse selection), we used the
COSSC 64-ms count rates and background fits
to determine where each of our pulses began and ended.  
To do this, we developed an interactive IDL routine to fit 
the Norris et al. (1996) pulse profile to the individual pulses within these bursts. 
The pulse profile function for the count rate C(t) can be written 
\begin{equation}
\rm{ C(t) =  A~exp~ \left( - {\left| \frac{t-t_{max}}{\sigma_{r,f}} \right|}^{\nu} \right)}
\end{equation}
where $\rm{t_{max}}$ is the time of maximum count rate, $\sigma_{\rm{r}}$ and 
$\sigma_{\rm{d}}$ are the count rate rise and decay time constants, and 
$\nu$ is the pulse ``peakedness''.  For an exponential rise and decay, $\nu = 1$.
When $\nu = 2$ and $\sigma_{\rm{r}} = \sigma_{\rm{d}}$ this shape describes a Gaussian.

With 64-ms resolution, we found that in many bursts pulses overlapped in a fashion
making them too complex
for us to fit individual pulses.  Other bursts
contained pulses which could be resolved, but none of their pulses
lasted long enough to span at least 4 time bins with SNR $\sim45$.  For
13 of the bursts, no processed 64-ms data was available.  We discarded bursts
which fell into any of these three categories.  This left us with
$\SelectedBursts$ bursts.
This is comparable to the 34 multi-pulse bursts used in LK96,
which included several pulses which appear to be overlapping when 
confronted with the 64-ms data.  To avoid overlap in our own analysis,
we used only bins which were dominated by a single pulse 
(at least $\CountPercent$ of the counts from one pulse).  
Within our $\SelectedBursts$ bursts, 
we identified $\SelectedPulses$ regions 
composed of at least 4 time bins dominated by a single pulse.  
The time bins selected for each pulse were consecutive in all but 
two cases (BATSE triggers 451 and 3290).  The 64-ms data for each of these
two bursts suggests that a short pulse occured near the middle of a longer
pulse, which forced us to fit two separate regions with a single decay
law. 

Our next step was
to test the $\Epk$-fluence relations (Equations 1 and 3) with each of the selected pulses.
Our motivation for emphasizing the   
$\Epk$-energy fluence relation (Eq. 3) as opposed the 
$\Epk$-photon fluence relation (Eq. 1) is that we believe 
that the former represents
a more physical quantity. 
It is possible (perhaps even likely) that the observed BATSE photon 
fluence is a poor representation of the bolometric photon fluence.  
The BATSE LAD energy window was
designed to contain the peak of GRB
energy spectra, \emph{not} the peak of the photon 
spectra.  By using 
energy fluence in place of photon fluence, we can avoid
the shakier assumption that the BATSE LAD
photon flux is proportional to the bolometric photon flux.
LK96 had attempted this but found that statistical 
errors in $\mathcal{E}$ were too large to be useful.
This was a result of their fixing $\alpha$ and $\beta$ in the spectral fitting.  When we fit 
the time-resolved spectra with variable $\alpha$ and $\beta$, we obtained
much smaller 
errors for $\mathcal{E}$, which made testing the $\Epk-\mathcal{E}$ relation
possible.  Nevertheless, we also fit Eq. 1 to our pulses
in this paper both for 
historical reasons and as a test of our interpretation.

\section{Results}

We fit both the $\Epk-\Phi$ and 
the $\Epk-\mathcal{E}$ relations to our $\SelectedPulses$ clean pulses   
using FITEXY (\cite{pres92}). 
Table 1 summarizes the results for each of the pulses in our sample.  The
first column is the BATSE trigger number.  The second column is the burst name,
which is also the date the burst triggered in the format YYMMDD.
The third column is the number of the LAD which was used for processing.
The fourth column lists the $\rm{t_{max}}$ from the Norris function fit to the pulse.
The fifth column 
is the energy fluence within the bins selected for fitting
in units $\rm{MeV~cm^{-2}}$.
The sixth column gives the number of bins selected for fitting.  
The seventh column is the fitted value $\Phi_{0}^{\rm{LK}}$ for
each pulse defined in Eq. 1.  The eighth and ninth columns are
the fitted values of $\Phi_{0}$ and $\rm{E_{pk(0)}}$ for each pulse as
defined in Eq. 3.  
Of course, $\Phi_{0}$
will only equal $\Phi_{0}^{\rm{LK}}$ if $\Epk \rm{F}_{N} = \rm{F}_{E}$.
Since the latter is not strictly true, we find that $\Phi_{0} \approx
\Phi_{0}^{\rm{LK}}$.  
For completeness, we also show the plots
of $\Epk$ versus $\mathcal{E}$ and their fits in Figure \ref{Figure2a}.

From the $\chi^{2}$ and the
number of fluence bins for each decay fit, we calculated the probability Q of
randomly getting a higher $\chi^{2}$ by chance.  Thus, $\rm{Q} \geq \sim 0.5$
represents very good fits, while $\rm{Q} \sim 0$ represents poor fits.
The Q values from fits of Eq. 3 to our pulses appear in the tenth column of Table 1.
If $\Epk$ does indeed cool linearly with $\mathcal{E}$
in all pulses selected for fitting, 
then when plotting the cumulative distribution of Q values, we would expect 
10\% of the pulses to have a Q less that 0.1, 20\% of the pulses to 
have a Q less than 0.2, and so on.  
Figure \ref{Figure3} shows the cumulative distribution of Q values for our pulses
with acceptable fits.  An excess of 
pulses with very high Q values would suggest a biased pulse selection process. 
A Kolmogorov-Smirnov test (P = $\PfromQfig$) 
applied to our distribution of Q values suggests that the set of $\SelectedPulses$ 
pulses is not too biased and roughly follows the distribution we would expect if all
of them are consistent with a linear decay of $\Epk$ with respect to energy fluence.  

By fitting the $\Epk-fluence$ law to the full observable duration of each
pulse and not just the flux decay phase, we could characterize our pulses
as ``hard-to-soft''. None of our pulses required 
a ``tracking'' classification, though many of the ambiguous pulses excluded
from this study
could be ``hard-to-soft'' or ``tracking''.
Three of the pulses in our sample (BATSE triggers 2316, 3491, and 3870) contain
pulses with negative values of $\Phi_{0}$.  However, all three of these pulses 
are still consistent with a positive value of $\Phi_{0}$. 
We remind the reader here that 
large absolute values of $\Phi_{0}$ (like those in these three pulses)
correspond to pulses with very little change
in $\Epk$, where $\frac{1}{\Phi_{0}} \equiv \frac{d \Epk}{d \mathcal{E}} \approx 0$. 
In such cases, small statistical errors in $\frac{d \Epk}{d \mathcal{E}}$ translate
to very large statistical errors in $\Phi_{0}$.  Even if
all pulses decay monotonically from hard-to-soft, we should expect to see a few
pulses to have negative values of $\Phi_{0}$.  Since all of our pulses are
consistent with a monotonic decay in $\Epk$, we adopt the hypothesis that all
pulses behave this way for the remainder of this paper and drop these three
pulses from our sample to simplify our calculations.  

\subsection{Distribution of $\Phi_{0}$}

The distribution of fitted $\Phi_{0}$ values appears in Figure  
\ref{Figure4}.
It is roughly log-normal where the mean of 
$\rm{log_{10}\Phi_{0}}$
is $\AverageLogPhi$ and the FWHM of $\rm{log_{10}~\Phi}$ is 
$\FWHMLogPhi$.
This distribution likely suffers some selection effects.  This
becomes obvious when one realizes that
$\Phi_{0} \approx - \frac{\Delta \mathcal{E}}{\Delta \Epk}$. We see that the
smallest absolute value of $\Phi_{0}$ is limited by the minimum energy fluence
which allows one to fit spectra (about $1~\rm{MeV~cm^{-2}}$ from Table 1)
and the energy window of BATSE (max $\Delta \Epk \approx 1870~keV$).  
There are no such limitations on the high side of this distribution,
since $\mid \Delta \Epk \mid$ can be arbitrarily small and
$\Delta \mathcal{E}$ is only limited by nature.  

\subsection{Testing the Invariance of $\Phi_{0}$ Among Pulses within Bursts}

LK96 reported that the decay constant
$\Phi_{\rm{0}}^{\rm{LK}}$ sometimes remains fixed from pulse to pulse
within some bursts.  Such behavior would
hint at a regenerative source rather than a  
single catastrophic event (such as \cite{mesz93}). 
However, the intrinsically narrow distribution of decay constants 
mentioned above and the relatively large confidence regions for each pulse's
value of $\Phi_{0}$ suggest
that many bursts would appear to have an invariant decay constant merely
by chance.  

As done earlier with a larger, but less reliable, set of pulses (\cite{crid98b}), we
calculated three statistics for each multi-pulse burst to test the invariance
of the $\Epk-fluence$ decay constant.  
We compared two of each bursts' M pulses at a time using the statistic
\begin{equation}
        \rm{X}_{ij}^{2} = \frac{[\Phi_{0}(i) - \Phi_{0}(j)]^{2}}
                          {\sigma_{\Phi_{0}(i)}^{2} +
\sigma_{\Phi_{0}(j)}^{2}}
\end{equation}
and then distilled the comparisons within each burst into a single
statistic to represent that burst.  These statistics are defined in 
Table 2.  Each is tailored for different
null hypotheses.
The statistic $G_{1}$ tests if \emph{at least} two pulses
in a burst are similar (and thus ``invariant''), while $G_{2}$ tests
if \emph{all} the pulses
have a similar decay constant.  $G_{3}$
tests for either a single good pairing or
several moderately close pairings.  We believe that this last statistic
is the most reasonable for testing our results since it does not
require that \emph{all} pulses decay at the same rate (as $G_{2}$ does) but
also does not discard information about multiple pulses repeating (as
$G_{1}$ does).
Finally, we calculated a table of probabilities P$(G,\rm{M})$
for our goodness-of-fit statistics $G$ based on 
a simple Monte Carlo simulation.  We 
created synthetic bursts with pulse decay parameters randomly sampled
from the observed distributions of $\Phi_{0}$ and $\sigma_{\Phi_0} / {\Phi_0}$.
To avoid any bias that intrinsic invariances would have on these
distributions, we created them using only one pulse from each burst.
The sample of bursts in this study has 
fewer bursts than previous works, and hence has fewer bursts with more than
one pulse.  The three versions of the $G$ statistic defined above are equivalent
when only two pulses appear in a burst.  Thus for this sample, with only 3 of the 
9 multipulse bursts having more than 2 pulses, these statistics are nearly equivalent.
The high probability that the 
observed repetitions occurred by coincidence leads us to conclude
that pulse decays are not invariant from pulse to pulse within bursts. 
Instead, we suggest that the distribution of $\Phi_{0}$ values seen in 
all bursts is narrow enough that an apparent invariance of $\Phi_{0}$ 
is inevitable in some burst.  We came to the same conclusion when examining 
$\Phi_{0}^{\rm{LK}}$ (\cite{crid98b}).

\section{Discussion}

Out of the $\SelectedBursts$ bursts to 
which we could fit a time-evolving Band GRB 
function, all contain at least one pulse consistent with
a linear decay of $\Epk$ with respect to energy fluence.
Of the $\SelectedPulses$ pulses in these bursts, all 
are consistent ($\rm{Q} > 0.001$; \cite{pres92}) 
with this decay pattern.  This is also
true when we fit the LK96 exponential decay of 
$\Epk$ with respect to photon fluence.
Besides LK96, other quantitative spectral evolution trends have been reported for GRBs.
The averaged temporal and spectral evolution for 32 bright GRBs has been 
calculated in \cite{feni98}.  The averaged photon flux evolution can be
described as both rising and decaying linearly with time.  The hardness, 
as measured
by $\Epk$ with $\alpha$ and $\beta$ held fixed, also appears to decay linearly with time
during the averaged burst ($\Epk = \rm{E_{pk(0)}} (1 - t/t_{0})$).  
This is clearly not representative of all bursts since the evolution in bursts of 
$\Epk$ is often complex (\cite{ford95,lian96}).  
These trends possibly reflect the physics dictating the envelope of emission. 
The fact that LK96 found the $\Epk$-fluence trend in many mingled pulses
may result from the fact that the burst envelope also evolves in this manner.

Since the hardness of this envelope appears to decay more slowly than the hardness 
during the pulses we observe, we might not expect to see this trend in our pulses. 
However, the degree of confidence of $\Epk$ in our fits, coupled
with the fact that energy fluence is often linear in time, makes the observations
of many bursts possibly consistent ($\rm{Q} > 0.001$) with this decay law.
Testing the distribution of Q values as we did in Fig \ref{Figure3}, we find
a probability P=0.001 that the pulses are realizations of linear Epk-time trend,
compared to P=$\PfromQfig$ for the linear Epk-energy fluence trend.
While the linear $\Epk$-time relation
does not seem to describe individual pulses as well as the $\Epk$-fluence relation,  the results are not conclusive.  

More pulses are clearly needed if one is to discriminate between
any two time-dependent spectral functions.  
One could simply wait for  bursts to occur or for a more sensitive instrument 
to be built.  However, it may
be possible to increase the number of fittable pulses using the existing BATSE
database.  Fitting a time-dependent spectral function directly to higher
time resolution data (or time-tagged event data) greatly reduces the number
of required fit parameters.  Another approach may be to analytically integrate
the time-dependent spectral function and fit that to integrated spectra, 
as has been done by Ryde \& Svensson (1998).  
By increasing the number of pulses,
it will become possible to make more definitive statements 
about the evolution of
prompt GRB emission and how it relates to the GRB afterglow.

\acknowledgements

AC thanks NASA-MSFC for the Graduate Student Research Program
fellowship.  It is also a pleasure to mention very useful discussions with
Ed Fenimore, Charles Dermer, Markus B\"{o}ttcher, and Rodney Hurley.  
This work is supported by NASA grant NAG5-3824.

\clearpage
\figcaption [FigMix.ps] {
The time history of BATSE trigger 543 seen both in
(a) the lower time resolution which allowed us to fit a spectrum to each bin  
and (b) 64-ms resolution.  Count rate is marked
in each plot as a histogram and $\Epk$, as determined from the data in the 
upper plot, is marked on both plots for convenience. 
While it is clear 
in the lower plot that there are at least 4 pulses within this burst,
only two peaks in flux are evident in the upper plot.  The fits of the
Norris function (Eq. 5) to these pulses are plotted here as dotted lines.
\label{Figure1}}

\figcaption [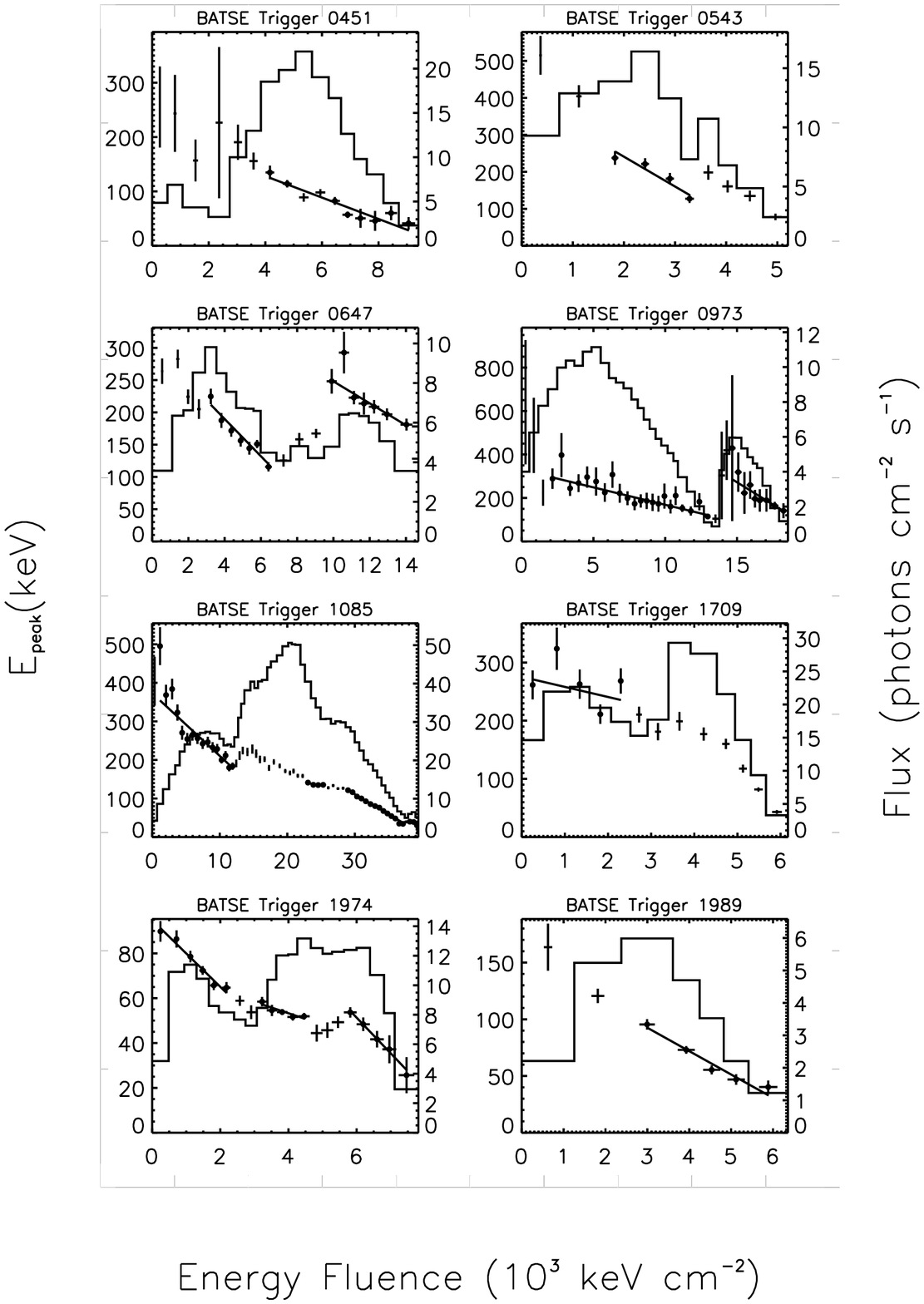] {The $\Epk$ (crosses) and photon flux (histogram)
evolution of 26 gamma-ray bursts with respect to energy fluence.  Bins used
to calculate the decay rate are marked with solid circles.  The best fit decay 
law also appears.  One sigma confidence bars for $\Epk$ and
energy fluence are shown for each bin.
\label{Figure2a}}

\figcaption [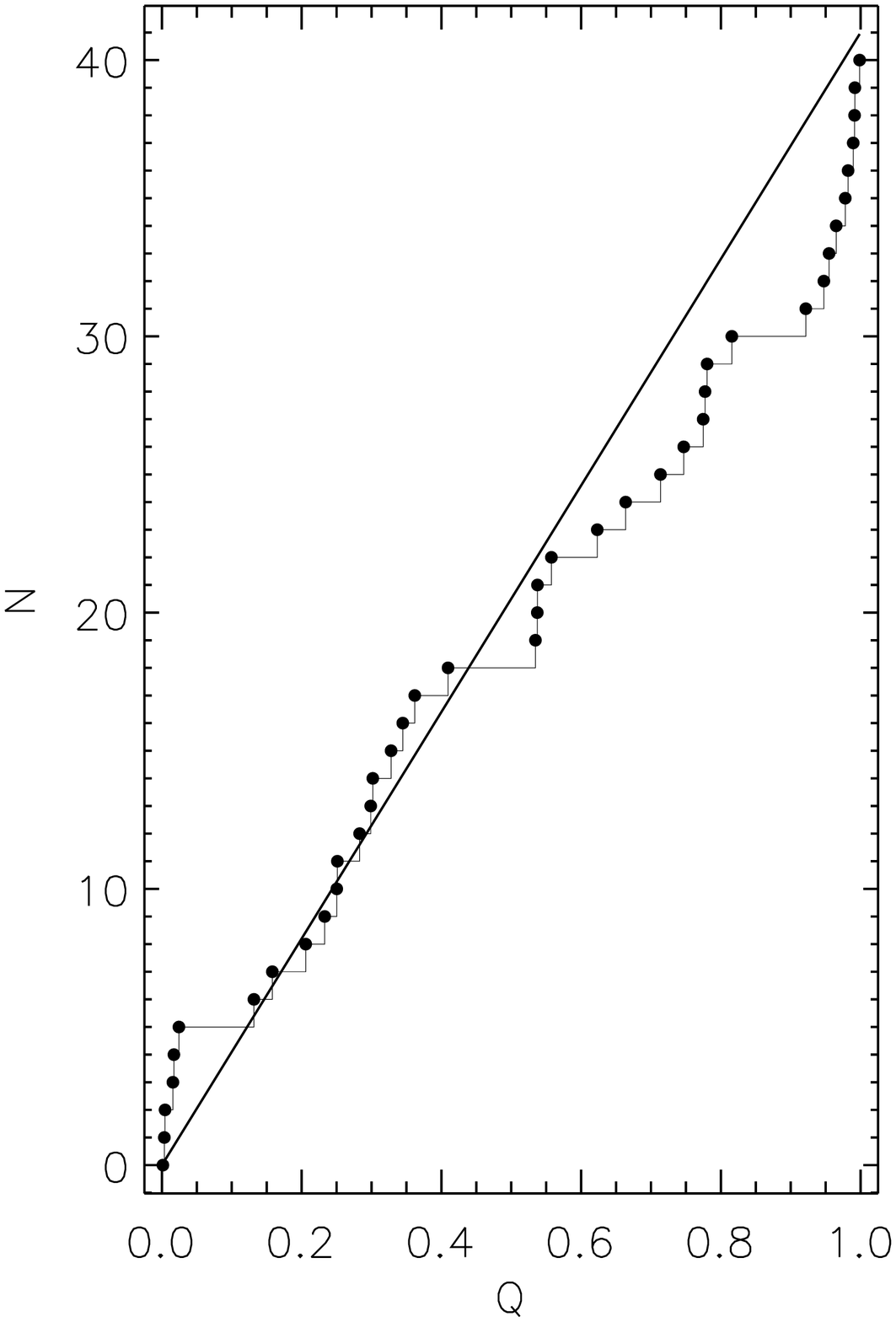] {Cumulative distribution of Q values for $\GoodPulses$ pulses.
In this plot, N is the fraction of pulses that have a Q value less than that of 
a certain pulse. 
If all of these GRB pulses cooled linearly as a
function of energy fluence, one would expect the cumulative Q distribution
to match the unit distribution (N=Q). 
The Kolmogorov-Smirnov test gives a probability
P = $\PfromQfig$ that this distribution is drawn from
the unit distribution.  From this, we conclude that our $\Epk-\mathcal{E}$ function 
(Eq. 3) adequately describes the pulses in this subset.
\label{Figure3}}

\figcaption [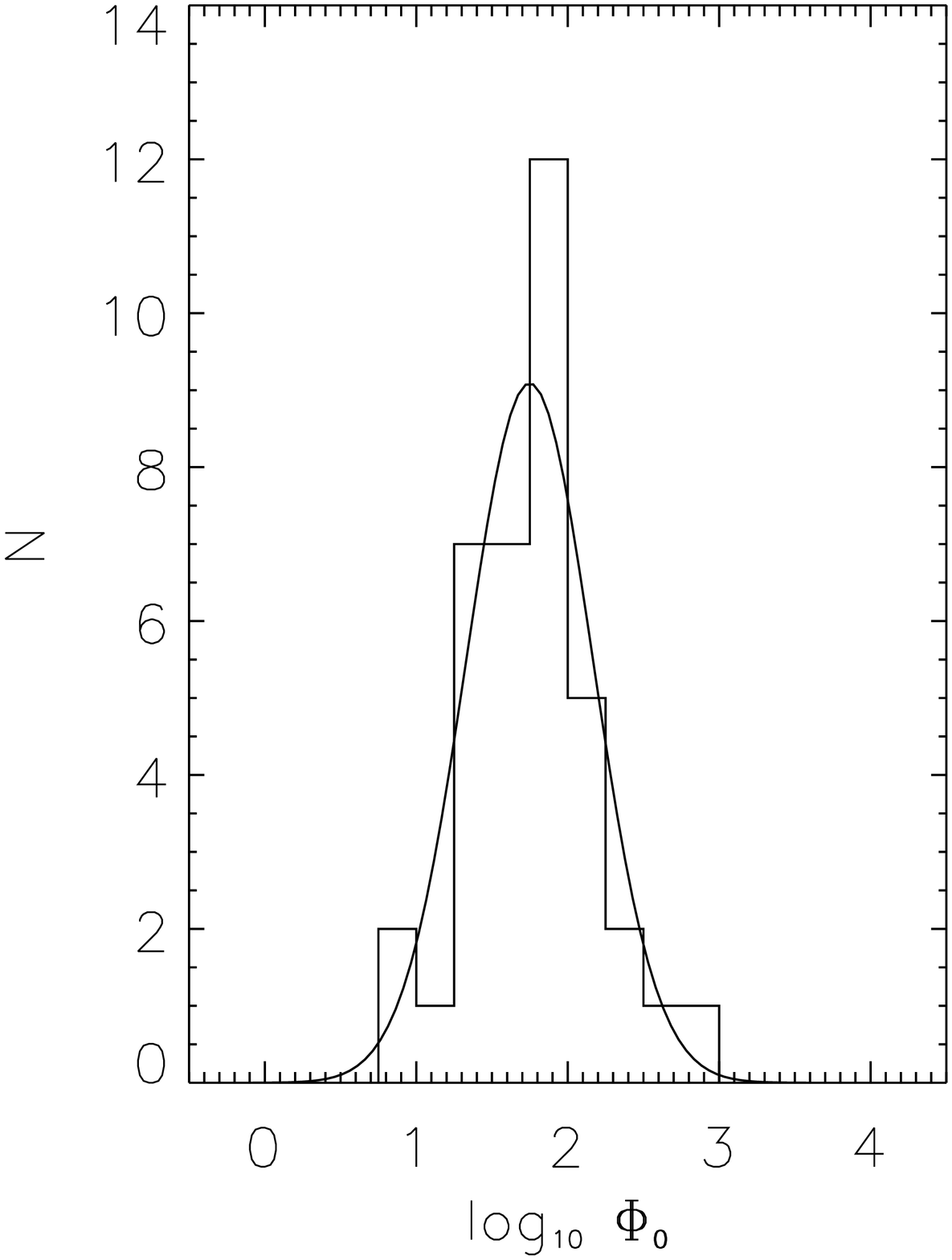] {This is the distribution of $\Phi_{0}$ values based on
$\GoodPulses$ pulses selected from $\SelectedBursts$ bursts.  In this plot, we show both a
histogram of values and the best-fit Gaussian distribution.
The distribution of $\Phi_{0}$ values is roughly log-normal where the mean of
$\rm{log_{10}\Phi}$ is $\AverageLogPhi$ and the FWHM of $\rm{log_{10}\Phi}$ is 
$\FWHMLogPhi$.  \label{Figure4}}

\clearpage

\begin{table} [h]
\tablenum{2}
\caption{The probabilities of getting such good values for each of our
three goodness-of-fit statistics $G$.
These probabilities are high enough to suggest that any
invariance seen in the data is purely coincidental.}
\vspace{10pt}
\label{table2}
\begin{center}
\begin{tabular}{lc}
Definition of $G_{\alpha}$ & P($G_{\alpha(\rm{random})} 
< G_{\alpha(\rm{observed})}$) \\
\hline
\hline
\\
$G_{1} \equiv \rm{min}~X_{ij}^{2}$ & $\PminX$ \\
\\
$G_{2} \equiv \rm{\sum_{i=1}^{M-1}\sum_{j=i+1}^{M}X_{ij}^{2}}$ & $\PsumX$ \\
\\
$G_{3} \equiv \rm{\prod_{i=1}^{M-1}\prod_{j=i+1}^{M}X_{ij}^{2}}$ &  $\PprodX$ \\
\\
\end{tabular}
\end{center}
\end{table}

%
%
%

\end{document}